\documentclass{elsart}

\usepackage{amssymb}
\usepackage{graphicx}
\usepackage{booktabs}
\begin{document}

\begin{frontmatter}

\title{Statistical Global Modeling of $\beta^-$-Decay Halflives Systematics Using Multilayer Feedforward Neural Networks and Support Vector Machines}
\thanks[talk]{\small URL: www.pythaim.phys.uoa.gr, E-mail: pythaim@phys.uoa.gr}

\author[1]{\normalsize N.~Costiris,}
\author[1]{\normalsize E.~Mavrommatis,}
\author[2]{\normalsize K.~A.~Gernoth,}
\author[3]{\normalsize J.~W.~Clark,}
\author[4]{\normalsize  and H.~Li}
\address[1]{Physics Department, Section of Nuclear and
Particle Physics\\
University of Athens, GR-15771 Athens, Greece}

\address[2]{
Institut f\"ur Theoretische Physik, Johannes-Kepler-Universit\"at,\\ 
A-4040 Linz, Austria\\
School of Physics and Astronomy, Schuster Building\\ The
University of Manchester, Manchester, M13 9PL, UK}

\address[3]{McDonnell Center for the Space Sciences and
Department of Physics\\ Washington University,
St.~Louis, MO 63130, USA}

\address[4]
{Dartmouth Medical School, Lebanon, NH 03756, USA}

\begin{abstract}

In this work, the beta-decay halflives problem is dealt as a nonlinear optimization problem, which is resolved in the statistical framework of Machine Learning (LM). Continuing past similar approaches, we have constructed sophisticated Artificial Neural Networks ($ANNs$) and Support Vector Regression Machines ($SVMs$) for each class with even-odd character in $Z$ and $N$ to global model the systematics of nuclei that decay 100\% by the $\beta^-$-mode in their ground states. The arising large-scale lifetime calculations generated by both types of machines are discussed and compared with each other, with the available experimental data, with previous results obtained with neural networks, as well as with estimates coming from traditional global nuclear models. Particular attention is paid on the estimates for exotic and halo nuclei and we focus to those nuclides that are involved in the r-process nucleosynthesis. It is found that statistical models based on LM can at least match or even surpass the predictive performance of the best conventional models of $\beta$-decay systematics and can complement the latter.

\end{abstract}

\end{frontmatter}

\section{Introduction}

Reliable,  quantitative estimates of $\beta^-$-decay halflives of nuclei far from stability are needed by the experimental exploration of the nuclear landscape at existing and future radioactive ion-beam facilities and by ongoing major efforts in astrophysics towards understanding of supernova explosions, and the processes of nucleosynthesis, notably the r-process~\cite{1} . In the nuclear chart there are spaces for some 6000 nuclides between the
$\beta$-stability line and the neutron-drip line. Although, $\beta^-$-decay properties have been already measured in terrestrial laboratories for some r-process key nuclei and more will be
measured in future facilities, the majority of $\beta$-decay rates of the involved neutron-rich nuclei should be estimated from theoretical models. Several approaches of different level of sophistication for determining $\beta$-halflives  have been proposed and applied over the years. One can mention the more phenomenological treatments based on Gross Theory ($GT$)~\cite{2}, as well as microscopic treatments that employ the $pn$ Quasiparticle Random-Phase Approximation ($pn-QRPA$) (in various versions) ~\cite{4},~\cite{10}  or the shell-model~\cite{13}. The latest hybrid version of the $RPA$ models developed by M\"oller and coworkers, combines the $pn-QRPA$ model with the statistical Gross Theory of  the first forbidden decay ($pn$QRPA+\textit{ff}GT)~\cite{15}. There are also some models in which the ground state of the parent nucleus is described by the extended Thomas-Fermi plus Strutinsky integral method, or the Hartree-Fock $BCS$ method, or other density functional method ($DF$) and which use the continuum $QRPA$ ($CQRPA$)~\cite{16}. Recently relativistic $pn-QRPA$ ($RQRPA$) models have been applied in the treatment of neutron-rich nuclei in the $N\sim50$, $N\sim82$ and $Z\sim28$ and $50$ regions~\cite{20}. Despite continuing improvements the predictive power of these conventional ``theory-thick''  models is rather limited for $\beta^-$-decay halflives of nuclei that are mainly far from stability, with deviations from experiment of at least an order of magnitude and considerable sensitivity to quantities that are poorly known. 

The recent advances in Artificial Inteligence ($AI$) and especially in statistical learning theory or Machine Learning (LM), notably Artificial Neural Networks ($ANNs$) and Support Vector Machines ($SVMs$) provide, an alternative opportunity to develop statistical models of observables of different systems that exhibit significant power. These models are ``theory-thin'' and are driven mainly from the data. For example, in the case of nuclei, any nuclear observable $X$ can be viewed as a mapping from the proton and neutron numbers $Z$ and $N$ (($Z$,$N$)$\to X$). In $LM$ one attempts to approximate the mapping by using only a subset of the data for $X$ (training data). $LM$-based models have already been developed for several nuclear properties ~\cite{19-2}, including atomic masses and ground state spins and parities. In this work, which continues previous studies in statistical modeling of nuclear halflife systematics [10-12], we present global models for the halflives of nuclear ground states that decay 100\% by the $\beta^-$ mode ($T_{\beta}$) using the $ANN$ and $SVM$ algorithms aiming to compare the two algorithms and to investigate further the potentiality of the modeling using learning machines. In Section 2 we briefly present the elements of the models. In Section 3 some of the results of the models are presented and evaluated and in Section 4 the conclusion and prospects of the present study are considered.

\section{The Models}

A Learning Machine consists of i) an input interface where, external, input variables to the device in coded form (for ex. $Z$ and $N$) are fed , ii) a system of intermediate elements or units that process the input, and iii) an output interface where an estimate of the corresponding observable of interest (say the beta halflife $T_{\beta}$) appears for decoding. Given an adequate body of training data (learning set), a suitable training algorithm is used to adjust the parameters of the machine to produce good performance on this set and good generalization  on test examples (test set) absent from the training set. The machine interpolates or extrapolates. $ANNs$ are nonlinear computational structures, inspired by biological neural systems, which consist of interconnected group (layers) of artificial neurons (processing units). The connections of the units (weights) determine their function~\cite{26}.  In particular, feedforward networks (multilayer perceptrons) on which we focus in this work although fully-connected have no lateral and feedback connections and the information flows from the input to the output. 
$SVMs$ are learning systems having a rigorous basis in the statistical learning theory developed
by Vapnick and Chervonenkis~\cite{27} (VC theory) and belong
to the class of Kernel methods (meaning that they implicitly perform a nonlinear mapping of the input data into a high-dimentional feature space). There are similarities as well as differences between $ANNs$ and $SVMs$. The differences have mainly to do with the tradeoff between complexity and generalization ability and with the ``nature'' of their parameters. During $SVM$ training a reduced number of training patterns (support vectors) is picked and determines the architecture (optimal number of neurons).

In this work we have constructed four separate $ANN$ and $SVM$ models that determine the halflives of the nuclides according to the pairing of $Z$ and $N$ (even-$Z$-even-$N$ ($EE$), even-$Z$-odd-$N$ ($EO$), odd-$Z$-even-$N$ ($OE$), and odd-$Z$-odd-$N$ ($OO$) ). We briefly list below the main features of these models and further information on the methodology is found in Ref. [9, 15] and [11, 16] for $ANN$ and $SVM$ models respectively. The four $ANN$ models are fully-connected, multilayer feedforward networks with architecture symbolized by $[2-5-5-5-1|81]$.  The activation function of the neuron-like units is given by hyperbolic-tangent sigmoid function in the intermediate (hidden) layers and a saturated linear function in the output layer. The two input units encode $Z$ and $N$ and the single output unit $\log_{10}T_{\beta}$. The Levenberg-Marquardt backpropagation algorithm has been used to train the network while implementing Bayesian regularization and cross-validation to improve generalization and using the Nguyen-Widrow method for initialization of the network. The  four $SVM$ models use a  $pA$-type Kernel (a composite kernel obeying Mercer's theorem and constructed linearly by combining the Anova ($A$) kernel  and a polynomial ($p$) kernel) with the two $pA$ parameters $p$ and $d$ equal to 1 and 8 respectively and the control parameter $C$ equal to $3.861 \times 10^?3$  for all four classes.

The experimental data used in developing our
models of $\beta$-decay systematics have been taken
from the $Nubase2003$ evaluation~\cite{22}.  Restricting attention to those
cases in which the ground state of the parent decays 100\%
through the $\beta^-$ channel, we form a subset of the
beta-decay data denoted by NuSet-A, consisting of 905
nuclides. We also formed a more restricted data set, called NuSet-B, by eliminating
from NuSet-A those nuclei having halflife greater than $10^6\,{ s}$. The halflives in this subset range from $0.15 \times 10^{-2}\,{ s}$ for $^{35}${\rm Na} to $0.20\times 10^{6}\,{ s}$ for $^{247}${\rm Pu}.  
NuSet-B consists of 838 nuclides (\textit{Overall set}):  672 ($80\%$) (\textit{learning set}) of them have been randomly chosen to train the $ANNs$ or to find the support vectors in the case of $SVMs$; of those left  83 ($10\%$) (\textit{validation set})  have been similarly chosen to validate the learning procedure in the case of $ANNs$ or to guide the determination of small number of parameter, mainly entering the inner-product Kernel in the case of $SVMs$; the rest  83 ($\sim10\%$) (\textit{test set}) have been similarly chosen to evaluate the accuracy of the prediction. Considering performance on the above sets we speak of operation in overall, learning, validation and prediction modes. Having excluded the few long-lived examples from NuSet-A, one is then dealing with a more homogeneous collection of nuclides, a property that facilitates the training of network models.  Accordingly, we have focused our efforts on NuSet-B, and since the examples still span 9 orders of magnitude in $T_{\beta}$, it is natural to work with the $\log_{10}T_{\beta}$.

\section{Results And Discussion}

The performance of our $ANN$ and $SVM$ global models is first evaluated in 
Table 1 by direct comparison with the experimental data using a commonly known statistical metric, namely the \textit{Root Mean Square Error} ($\sigma_{\rm RMSE}$):
\begin{equation}
{\sigma_{\rm RMSE}} = \left[ {\frac{1}{N}\sum_{i=1}^n {\left( {y_i  - \hat y_i } \right)^2 } } \right]^{1/2}.
 \label{eq:1}
\end{equation}
where $y_i \equiv \log_{10}T_{\beta,{\rm calc}}$ and $\hat{y}_i \equiv \log_{10}T_{\beta,{\rm exp}}$
and $N$ the total number of nuclides in each case. We mention for comparison the $\sigma_{\rm RMSE}$ values of an earlier $ANN$ model~\cite{19-4} which equal to  1.08 and 1.82 for total learning and test sets, respectively, as well as those of the $ANN$ model developed recently by means of the whole basis  ~\cite{19-1}, which equal to 0.53, 0.60 and 0.65 for the total learning, validation and test sets respectively.
In Table 2, a further assessment is made by tabulating the performance measures of M\"{o}ller and collaborators~\cite{15}: $M$, its standard deviation  $\sigma_{M}$ and $\Sigma$ . These measures are defined as follows in terms of the variable $r_i = y_i /{\hat y}_i$:
\begin{equation}
M  = \frac{1}{N}\sum_{i=1}^N {r_i }, \quad
\sigma_M  = \left[ {\frac{1}{N}\sum_{i=1}^N{\left( {r_i  - M } \right)^2 } } \right]^{{1 \mathord{\left/
 {\vphantom {1 2}} \right.
 \kern-\nulldelimiterspace} 2}}, \quad
 \Sigma  = \left[ \frac{1}{N}\sum_{i=1}^N  r_i^2   \right]^{{1 \mathord{\left/
 {\vphantom {1 2}} \right.
 \kern-\nulldelimiterspace} 2}}.\quad
 \label{eq:2}
\end{equation}
Superior models should have   $M$, $\sigma_M$  and $\Sigma$ near zero.
A comparison follows with two recent global theory-thick models of the above collaboration, namely the $FRDM+pnQRPA$ and the $pnQRPA+\textit{ff}GT$ models.~\cite{15}. Finally, in Fig. 1, halflives of $\beta^-$-decaying nuclides that
are found near or on a typical r-process path with neutron separation energy
below 3 MeV derived by means of the present $LM$-models are  compared with those from $pn$QRPA+\textit{ff}GT calculation and a calculation by Pfeiffer and coworkers [18] (labeled GT*) based on the early Gross Theory (GT) of Takahashi et al. [2a] with updated mass values.

From the above comparison one can conclude that $LM$-based models give similar results, which are close to experimental data. In Fiq. 1, the results  given by the $SVM$ model are almost equal with the experimental values. This occurs  because, unlike neuron network, where the connections between neurons are random values, SVMs use exclusively only almost all
training patterns as ``nodes'',  a fact that may lead to better interpolation but often worse extrapolation. Furthermore, the comparison of the results derived by the $LM$-modes with those of the models presented in Table 2 as well with others [15,16] lead to the conclusion that the former perform equally or better than the later. This is partially ascribed to the larger number of parameters of the $LM$-based models. Regarding the performance of the present $ANN$ and $SVM$ models with respect to that of previous $LM$-based models the consideration of different quality measures slightly favors the present ones. However, a more detailed analysis (see Ref. [15]) shows that the subdivision of the data into four ($Z$,$N$) parity classes can lead to spurious fluctuations in the prediction of lifetimes for nuclides of isotopic and isotonic chains. This favors the use of the $ANN$ model developed recently by means of the whole basis~\cite{19-1}.  

\begin{table}[htdp]
\caption{\textit{Root-mean-square errors} $\sigma$ ($\equiv{\sigma_{\rm RMSE}}$) (Eq. ~\ref{eq:1}) for learning, validation and test sets, achieved by the current $ANN$ and $SVM$ models (developed for even-even, even-odd, odd-even and odd-odd classes, in $Z$ and $N$) of $\beta^-$-decay halflives (with a cutoff at $10^6s$).}

\begin{center}
    \begin{tabular}{ccccccccccccc}
           \hline

    &\multicolumn{6}{c}{$ANN$ models}&\multicolumn{6}{c}{$SVM$ models}\\
           \hline
&\multicolumn{2}{c}{Lear. Set}&\multicolumn{2}{c}{Val. Set}&\multicolumn{2}{c}{Test Set}&\multicolumn{2}{c}{Lear. Set}&\multicolumn{2}{c}{Val. Set}&\multicolumn{2}{c}{Test Set}\\
       \hline
   Class&$N$&$\sigma$&$N$&$\sigma$&$N$&$\sigma$&$N$&$\sigma$&$N$&$\sigma$&$N$&$\sigma$\\
    \hline
   \small EE&131&0.36&16&0.41&16&0.62&131&0.55&16&0.57&16&0.62\\
  \small EO&179&0.38&22&0.44&22&0.39&179&0.41&22&0.42&22&0.51\\
  \small OE&172&0.44&21&0.46&21&0.53&172&0.41&21&0.47&21&0.47\\
  \small OO&190&0.52&24&0.42&24&0.33&190&0.52&24&0.40&24&0.52\\
    \hline
  \bf Total& \bf 672& \bf 0.41& \bf 83& \bf 0.44& \bf 83& \bf 0.51& \bf 672& \bf 0.47& \bf 83& \bf 0.46& \bf 83& \bf 0.53\\
           \hline

 \end{tabular}
\end{center}

\label{default}
\end{table}

\begin{table}[p]
\caption{\small Quality indices  $M$,  $\sigma_{M}$ and $\Sigma$ (Eq.~ \ref{eq:2}) for the present $ANN$ models in Overall (a) and Prediction (b) Modes and for the FRDM+$pn$QRPA and $pn$QRPA+\textit{ff}GT models of Ref.~\cite{15}. The number $n$ stands for the nuclides with experimental halflives below the prescribed limit.}
\centering
\begin{tabular}{lllllllll}
\hline
\multicolumn{1}{c}{\small \textbf{T$_{\beta,exp}$}} & \multicolumn{4}{l}{ \textbf{\small (a) $ANNs$ - Overall Mode} } & \multicolumn{4}{l}{ \textbf{\small (b) $ANNs$ - Prediction Mode} }\\ 
\hline
\multicolumn{1}{c}{$(s)$} & \multicolumn{1}{c}{$n$} & \multicolumn{1}{c}{$M$} & \multicolumn{1}{c}{$\sigma_{M}$} &  \multicolumn{1}{c}{$\Sigma$}  & \multicolumn{1}{c}{$n$} & \multicolumn{1}{c}{$M$} & \multicolumn{1}{c}{$\sigma_{M}$} & \multicolumn{1}{c}{$\Sigma$} \\ 
\hline

\multicolumn{1}{l}{$<1$} & \multicolumn{1}{c}{252} & \multicolumn{1}{c}{0.02}  & \multicolumn{1}{c}{0.30} & \multicolumn{1}{c}{0.30}  & \multicolumn{1}{c}{8} & \multicolumn{1}{c}{-0.19} & \multicolumn{1}{c}{0.61} & \multicolumn{1}{c}{0.64} \\ 

\multicolumn{1}{l}{$<10$} & \multicolumn{1}{c}{396} & \multicolumn{1}{c}{0.02} & \multicolumn{1}{c}{0.35}  & \multicolumn{1}{c}{0.35}  & \multicolumn{1}{c}{28} & \multicolumn{1}{c}{-0.16}  & \multicolumn{1}{c}{0.57} & \multicolumn{1}{c}{0.60}\\ 

\multicolumn{1}{l}{$<100$} & \multicolumn{1}{c}{529} & \multicolumn{1}{c}{0.03}  & \multicolumn{1}{c}{0.36} & \multicolumn{1}{c}{0.36}  & \multicolumn{1}{c}{54} & \multicolumn{1}{c}{-0.05}  & \multicolumn{1}{c}{0.51} & \multicolumn{1}{c}{0.52} \\ 

\multicolumn{1}{l}{$<1000$} & \multicolumn{1}{c}{653} & \multicolumn{1}{c}{0.05} & \multicolumn{1}{c}{0.39}  & \multicolumn{1}{c}{0.39}  & \multicolumn{1}{c}{77} & \multicolumn{1}{c}{0.03}  & \multicolumn{1}{c}{0.49} & \multicolumn{1}{c}{0.49} \\ 

\multicolumn{1}{l}{$<10^6$} & \multicolumn{1}{c}{838} & \multicolumn{1}{c}{0.00} & \multicolumn{1}{c}{0.45} & \multicolumn{1}{c}{0.45}  & \multicolumn{1}{c}{83} & \multicolumn{1}{c}{0.03} & \multicolumn{1}{c}{0.52} & \multicolumn{1}{c}{0.53}  \\

\hline
\multicolumn{1}{c}{\small \textbf{T$_{\beta,exp}$}} & \multicolumn{4}{l}{ \textbf{\small (c) $FRDM+pnQRPA$}~\cite{15}} &  \multicolumn{4}{l}{ \textbf{\small (d) $pnQRPA+\textit{ff}GT$}~\cite{15} }\\ 
\hline
\multicolumn{1}{c}{$(s)$} & \multicolumn{1}{c}{$n$} & \multicolumn{1}{c}{$M$} & \multicolumn{1}{c}{$\sigma_{M}$} &  \multicolumn{1}{c}{$\Sigma$}  & \multicolumn{1}{c}{$n$} & \multicolumn{1}{c}{$M$} & \multicolumn{1}{c}{$\sigma_{M}$} & \multicolumn{1}{c}{$\Sigma$} \\ 
\hline

\multicolumn{1}{l}{$<1$} & \multicolumn{1}{c}{184} & \multicolumn{1}{c}{0.03} & \multicolumn{1}{c}{0.57}  & \multicolumn{1}{c}{0.57}  & \multicolumn{1}{c}{184} & \multicolumn{1}{c}{-0.08} & \multicolumn{1}{c}{0.48}  & \multicolumn{1}{c}{0.49} \\ 

\multicolumn{1}{l}{$<10$} & \multicolumn{1}{c}{306} & \multicolumn{1}{c}{0.14}  & \multicolumn{1}{c}{0.77}  & \multicolumn{1}{c}{0.78} & \multicolumn{1}{c}{306} & \multicolumn{1}{c}{-0.03} & \multicolumn{1}{c}{0.55} & \multicolumn{1}{c}{0.55}\\ 

\multicolumn{1}{l}{$<100$} & \multicolumn{1}{c}{431} & \multicolumn{1}{c}{0.19} & \multicolumn{1}{c}{0.94}  & \multicolumn{1}{c}{0.96} & \multicolumn{1}{c}{431} & \multicolumn{1}{c}{-0.04} & \multicolumn{1}{c}{0.61} & \multicolumn{1}{c}{0.61} \\ 

\multicolumn{1}{l}{$<1000$} & \multicolumn{1}{c}{546} & \multicolumn{1}{c}{0.34}  & \multicolumn{1}{c}{1.28}  & \multicolumn{1}{c}{1.33}  & \multicolumn{1}{c}{546} & \multicolumn{1}{c}{-0.04} & \multicolumn{1}{c}{0.68}  & \multicolumn{1}{c}{0.68} \\ 

\multicolumn{1}{l}{$<10^6$} & \multicolumn{1}{c}{$-$} & \multicolumn{1}{c}{$-$} & \multicolumn{1}{c}{$-$}  & \multicolumn{1}{c}{$-$}& \multicolumn{1}{c}{$-$} & \multicolumn{1}{c}{$-$} & \multicolumn{1}{c}{$-$} & \multicolumn{1}{c}{$-$} \\ 

\hline
\end{tabular}
\label{tab:mollerindicies}
\end{table}

\begin{figure}[hbt]
\centering
\includegraphics[width=3.7in]{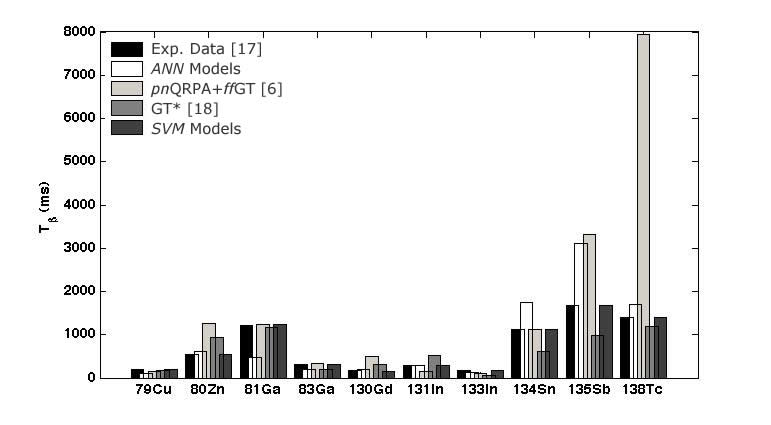}
\caption{\label{fig:fig11} Halflives for $\beta^-$-decaying nuclides that are found near or on a typical r-process path with the neutron separation energy lower or equal to 3 MeV derived by means of the present $ANN$ and $SVM$ models are compared with the experimental data and those from $pn$QRPA+\textit{ff}GT~\cite{15} and GT*~\cite{28}  calculations.}
\end{figure}

\section{Conclusion and Prospects}

In this work, the beta-decay halflives problem is dealt as a nonlinear optimization problem, which is resolved in the statistical framework of Machine Learning using Artificial Neural Networks ($ANNs$) and Support Vector Regression Machines ($SVMs$). It seems that both $ANNs$ and $SVMs$ demonstrate similar performance and that our statistical large-scale calculations can match or even surpass the predictive performance of the best conventional global calculations outside the stable valley. Moreover, this way of confrontation of the beta-halflives problem could give an estimation of  the degree that  the nucleonic numbers determine the beta-decay systematics of a nuclear system. Accordingly, we plan further studies of the systematics of beta decay using $LMs$, with the object of continued enhancement of their predictive power and of possible gaining of some new physical insight.

\section{Acknowledgments}

This research has been supported in part by the U.S. National Science Foundation under Grant No. PHY-0140316 and by the University of Athens under Grant No.70/4/3309.

\end{document}